# Mapping of Affymetrix probe sets to groups of transcripts using transcriptional networks

Michel Bellis[1,2*]

[1]CNRS,CRBM, UMR-5237, 1919 Route de Mende, 34293, Montpellier, France

[2]UMSF, UMR-5237, Montpellier, France

**ABSTRACT**

**Motivation:** Usefulness of analysis derived from Affymetrix microarrays depends largely upon the reliability of files describing the correspondence between probe sets, genes and transcripts. In particular, in case a gene is targeted by two probe sets, one must be able to assess if the corresponding signals measure a group of common transcripts or two groups of transcripts with little or no overlap.
**Results:** Probe sets that effectively target the same group of transcripts have specific properties in the trancriptional networks we constructed. We found indeed that such probe sets had a very low negative correlation, a high positive correlation and a similar neighbourhood. Taking advantage of these properties, we devised a test allowing to group probe sets which target the same group of transcripts in a particular network. By considering several networks, additional information concerning the frequency of these associations was obtained.
**Availability and Implementation:** The programs developed in Python (PSAWNpy) and in Matlab (PSAWNml) are freely available, and can be downloaded at http://code.google.com/p/arraymatic/. Tutorials and reference manuals are available at Bioinformatics online (Supplement 1) or from http://bns.crbm.cnrs.fr/softwares.html.
**Contact:** mbellis@crbm.cnrs.fr
**Supplementary information:** Supplementary data are available at http://bns.crbm.cnrs.fr/download.html.

## 1 INTRODUCTION

Since its introduction in the 90's, laboratories and companies involved in the development of microarray technology produced numerous different types of platforms. Affymetrix technology which is considered here in priority because it is the most frequently referenced platform, designed several chip models intended to quantify the expression level of transcripts by probing their 3' end (3'-IVT format). These chips display a complicated relationship between probes and genes. First, there exists redundancy at the probe level, and when a gene is used in a particular chip layout, it is always targeted by several different probes which form a group of 11 or 16 probes called a probe set. Second, a significative portion of probed genes are targeted by several probe sets, which is the consequence of using expressed sequence tags (EST) to design probe sets, at a date where the knowledge of genomic sequences was incomplete.

Soon after the first chips were released by the manufacturer, many authors noticed some inconsistancies either in probe set definitions (i.e. the set of probes considered as targeting the same gene(s) or transcript(s)) or in probe set descriptions (i.e. the information indicating which gene(s) or transcript(s) are targeted), and advocated for their improvement (Yu et al., 2007; Harbig et al., 2005). The main stream followed by researchers engaged in this effort consisted in using either genomic and/or transcript sequences to redefine the probe sets in order to optimize the quality of results by enforcing that newly defined probe sets target only one gene or even one transcript if alternative transcripts exist (Dai et al., 2005). The main drawback of this approach is that cdf files that combine the probes to define each probe set, and that are used by the algorithms which calculate the signal from raw data, must be redefined too. As several groups have independantly developed their own cdf file besides the official one delivered by Affymetrix, users can encounter difficulties in selecting the one which fits best to their needs. Another problem is the difficulty of using the results delivered with modified cdf files. For example, some specialized software may only use Affymetrix probe set names in the entry, and might not recognize proprietary probe set names. Along the same lines, comparing results obtained with different cdf files would be impossible. Finally, for those interested in massive analysis that rely on files stored in repositories like Gene Expression Omnibus (GEO, http://www.ncbi.nlm.nih.gov/geo/) or ExpressArray (http://www.ebi.ac.uk/arrayexpress/), it would be impossible to use the results that do not come with the raw files to recalculate signals with modified cdf files. For all these reasons, it seems sensible to maintain the original definition of probe sets as designed by Affymetrix, but to direct efforts toward improvement of their description, as already done in several published studies (Chalifa-Caspi, 2004 ; Ballester et al., 2010).

In this perspective, if two different probe sets are assumed to target the same gene(s) or transcript(s), it is of paramount importance to determine if such an assumption is true (Cui et al., 2009). Until now very few methods have been developed to answer this question, apart from the proposal of using sequence information, which does not allow to take a decision in particular circumstances. Actually, in absence of an exhaustive knowledge of all splice variants, it is not possible with this method to exclude completely that two probe sets located in different exons of the same gene do not target different transcripts. To circumvent this difficulty, it has been proposed to consider that two probe sets which are mapped within the same gene sequence, should be considered as targeting the same transcript(s) only if their signals measured in a large number of different biological conditions show a high correlation (Elbez et al., 2006). We have retained this idea, but propose to develop it in another direction. We have demonstrated in a previous publication that Pearson's correlation coefficient could not capture the complex patterns of correlation that are yielded when two probe sets are considered in the context of numerous different conditions (Hennetin et al., 2009). Instead, we have developed a new method to calculate both positive and negative correlation coefficients between a pair of probe sets by using the two strings which describe the direction of significative variations observed for each probe set in a large series of comparisons (e.g. string INNDIDi... which indicates that a given probe set i is increased, decreased or does not vary in the first and fifth, in the fourth and sixth, and in the second and third comparisons, respectively, would be used in conjunction with the corresponding string for probe set j, e.g.

IINDDDj..., to calculate positive and negative correlation coefficients between probe sets i and j). This method applied to all the possible pairs of probe sets results in two covariation matrices (one for positive and the other for negative correlation coefficients) that can be seen as a transcriptional network. These networks capture much more information than simple correlation calculated on signals, since they are able to take in account both positive and negative correlation between all probe sets, and we propose to use them to ascertain that two probe sets mapped to the same gene are to be considered as targeting the same transcript(s). Additionally, using several networks based on different sets of biological conditions, allowed us to construct several models each characterized by a given level of reproducibility.

The development of this new method we call PSAWN (for Probe Set Assignment With Networks) resulted in a program composed of two parts. One developed in Python (PSAWNpy) is intended to recover and to organize all the probe set information by interrogation of Ensembl (http://www.ensembl.org/index.html) and AceView (http://www.ncbi.nlm.nih.gov/IEB/Research/Acembly/, Thierry-Mieg and Thierry-Mieg, 2006) databases. Another developed in Matlab (PSAWNml) allows to find which probe sets target the same group of transcripts and to construct description files with different levels of reproducibility. We applied these programs to all the 3'-IVT Affymetrix chips in human, mouse and rat where it was possible to construct more than 10 networks and created a new resource allowing users to load description files corresponding to the most popular Affymetrix chips (http://bns.crbm.cnrs.fr/download.html).

## 2 METHODS

### 2.1 Construction of networks

Networks were constructed as indicated in Hennetin et al (2009), and are based on series of experiments (GSE) downloaded from GEO in September 2007. In order to take in account sampling effects that could bias network structure, we decided to randomly partition all biological conditions, that is couples of samples (GSM) belonging to the same experimental condition, in groups of 30 conditions, and to construct several networks by comparing any two groups of biological conditions, which generates each time 30x30=900 comparisons (a given group can be used in several comparisons and networks are therefore not fully independent). Table 1 lists the Affymetrix chips we used in this study, with the different names associated to them.

**Table 1. Chips used.**

| Affymetrix Name | GEO Name | Our Name |
|---|---|---|
| Human Genome U95A | GPL91 | m2 |
| Human Genome U133A | GPL96 | m3 |
| Murine Genome U74 Version 2 | GPL81 | m5 |
| Mouse Genome 430 2.0 | GPL1261 | m8 |
| Mouse Expression 430A | GPL339 | m27 |
| Rat Genome U34 | GPL85 | m6 |

GEO stands for Gene Expression Omnibus.

### 2.2 Recovering probe and probe set information

PSAWNpy is a Python program which allows to import user data registered in tabular format as well as public data stored in Ensembl and in AceView data bases into local Berkeley data bases as explained in the corresponding tutorial (Figures 1 to 10 in Supplement 1). Processing of these data allows to find correspondence between probes and genes, and to generate text tables indicating the probe sets that target each gene with at least one probe. These tables also display other information indicating the targeting probes, targeted and not targeted known transcripts, and location of probes (either in exons, in introns or upstream or downstream of the gene in a limit of 2 kb (Figure 11 and 12 of Supplement 1). As some probes are not located in known genes, we created a special category called GOP (group of probes, none of wich is separated from its direct neighbours by more than 2 kb).

### 2.3 Class and gene assignation of probe sets

Data generated by PSAWNpy are processed by PSAWNml, a Matlab program which constructs biclusters of probe sets and genes, and assigns them to different classes, according to the complexity of the relationship between the number of targeted genes and the number of targeting probe sets. Another task of PSAWNml is to study, within a series of networks, some characteristics of paired probe sets (i.e. probe sets targeting the same gene(s)), in order to set up a test to ascertain that given paired probe sets are similar (i.e. are assumed to target the same transcript(s)) (Figures 16 to 53 in Supplement 1). Further, all probe sets that are similar are collected together, thereby defining a group of probe sets that target a group of transcript(s), and enabling biclustering between probe sets and known and/or unknown transcripts. Finally, PSAWNml assigns each probe set to a single gene. When several genes are targeted, PSAWNml selects successively genes with the following characteristics until only one gene is left: the highest number of targeting probes, the best target type (probes in exons > probes in introns), the maximal ratio between the number of targeting probe sets and the number of groups of probe sets targeting the same transcript(s), the minimal number of targeting probe sets, the gene source (Ensembl> AceView>GOP), and finally the first gene in alphabetic order.

# 3 RESULTS

## 3.1 Definition of probe set classes

Customarily, probe sets are distributed according to the multiplicity of genes they target, which leads to a distinction between probe sets that target only one gene from those targeting several genes. However, the relationship between probe sets and genes can be considered as a bi-clustering problem, leading to a more complicated description.

Taking into account that a probe set can target either only one gene or several genes, and, symetrically, that a gene can be targeted either by a single probe set or by multiple probe sets, we could think *a priori* that effective relationships between probe sets and genes amounts to four classes the names of which could be abbreviated SS,SM,MS, and MM by indicating successively the number (single (S) or multiple (M)) of probe set(s) and of gene(s) in a particular bicluster. But practical considerations add another layer of complexity. In particular, construction of biclusters could be expected to depend on the application of a recursive algorithm, and we could then distinguish biclusters according to the depth of recursivity needed to reach the stop condition. In this perspective SS, SM and MS biclusters are constructed within one step (depth equal one), but biclusters with multiple probe sets and multiple genes may need one step (class MM), two steps (class CX, for complex) or more than two steps (class HX, for hyper-complex). As shown in figure 1, according to the number of recursive steps needed, the biclusters have contrasted densities of points: they are either saturated (classes SS (not depicted), SM, MS and MM), or more than 50% (class CX) or less than 30% (class HX) of their intersections corresponding to a relationship between a probe set and a gene (see also columns entitled Occ in Table 2). Contrary to classes MM and CX, class HX may contain biclusters of very large size.

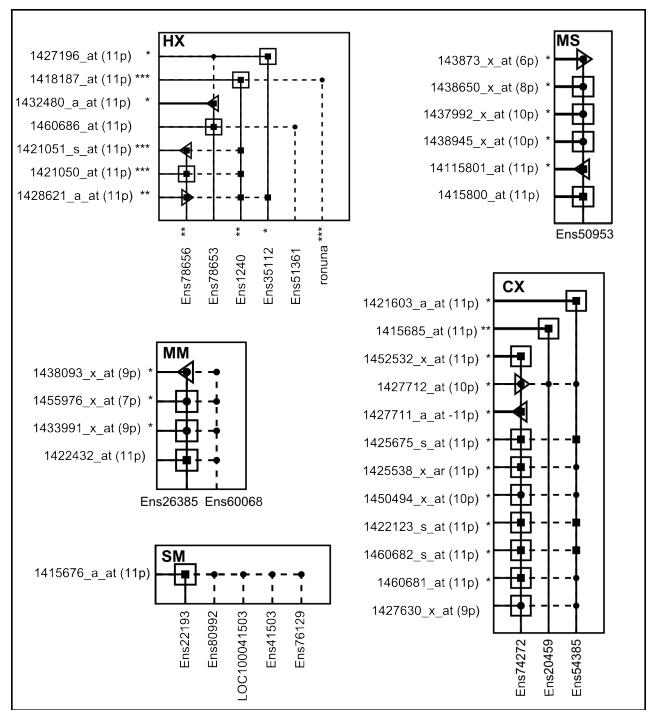

**Fig. 1. Classification of biclusters of genes and probe sets (m27-1p).** In each sketch exemplifying a particular class, the probe set name without asterisk is the one used to find genes and probe sets which are related to it. The number of asterisks indicates the depth at which the recursive algorithm finds the related gene(s) and probe set(s). A continuous line joins each probe set to the gene it is assigned to by PSAWNml (a larger horizontal line indicates a probe set which targets a single gene). Interrupted lines join probe sets to all other targeted but not assigned genes. Square and round symbols at the intersection between horizontal (probe set) and vertical (gene) lines indicate repectively that the gene is targeted by the maximum possible number of probes or by a smaller number of probes. Among genes that are not targeted by the maximum possible number of probes, a larger symbol distinguishes genes that are targeted by the greatest number of probes. In case several probe sets are assigned to the same gene, as in class MS, MM, CX and HX, similar probe sets , i.e. targeting the same transcript(s), are boxed by a common symbol (square, left or right triangle). Gene names starting with 'Ens' are shortened Ensembl IDs (Ens78653 corresponds to ENSMUG00000078653), and other names are AceView IDs

It must, however, be emphasized that classification depends greatly on the probe number limit, i.e. the minimal number of targeting probes required to consider that a probe set detects a gene. For example the bicluster constructed from 1415856_at in chip m27, with a probe number limit equal to one (m27-1p), belongs to class HX, and has 22350 filled intersections corresponding to 1943 probe sets and 10053 genes. But if we set the probe number limit to seven (m27-7p), the corresponding bicluster is classified differently as CX, and contains only 6 filled intersections corresponding to four probes and three genes. As shown in table 2, increasing the probe number limit from one to seven has a drastic influence on the distribution of biclusters, probe sets and genes in the different classes (SS and MS class sizes increase when other class sizes decrease). By considering only probe sets that target a single gene, in order to eliminate the effect of possible cross-hybridization with other targeted sequences, we observed that the number of similar probe sets was smaller than expected when one of the two probe sets target the common gene with less than seven probes (figure 29 of Supplement 1). In this category of probe sets, the probes which do not target the assigned gene, target *a priori* nothing else in the transcriptome, and we think that they are more subject to random hybridization, which blurs the signal. We therefore considered that a probe number limit set to seven maintained a good balance between specificity (if the probe number limit is too small, the number of genes considered as targets, even though they don't really participate in the signal, is too high) and sensitivity (if the probe number limit is too high, the number of genes discarded, even though they really participate in the signal, is too high).

### 3.2 PSAWN method

Column M of Table 2 shows that up to 77% of probe sets of a chip are assigned to a gene that is targeted by at least one another probe set (on all chips, with a probe number limit equal to 7, the average is 53%). Such a high proportion of multiple probe sets requires a method to process this type of probe sets in order to know if they target identical transcripts. We devised an original method based on caracteristic properties of these multiple probe set in networks to adress this question.

**Table 2. Biclusters, probe sets and gene frequency.**

| | | | | No | SS | SM | | MS | | MM | | | CX | | | HX | | | | M | |
|---|---|---|---|---|---|---|---|---|---|---|---|---|---|---|---|---|---|---|---|---|---|
| Sp | Chi | NetN | PN | P | BPG | BP | G | BG | P | B | P | G | B | P | G | Oc | B | P | G | Oc | P | %T |
| Hs | m2 | 21 | 1 | 127 | 4471 | 1535 | 4170 | 1034 | 2447 | 134 | 294 | 295 | 869 | 2586 | 4273 | 65 | 22 | 1149 | 6119 | 21 | 6476 | 51 |
| | | | 7 | 271 | 5537 | 1105 | 2785 | 1319 | 3104 | 130 | 277 | 282 | 692 | 2188 | 3452 | 66 | 13 | 127 | 403 | 31 | 5696 | 45 |
| | m3 | 18 | 1 | 503 | 5117 | 1010 | 2719 | 2132 | 5346 | 173 | 406 | 382 | 1127 | 3770 | 4324 | 65 | 11 | 6131 | 19473 | 25 | 15653 | 70 |
| | | | 7 | 1125 | 7366 | 594 | 1447 | 3612 | 9268 | 161 | 402 | 349 | 970 | 3420 | 3135 | 66 | 9 | 108 | 116 | 29 | 13198 | 59 |
| Mm | m5 | 35 | 1 | 1278 | 4680 | 1306 | 3692 | 871 | 1912 | 112 | 235 | 253 | 721 | 1991 | 3354 | 66 | 8 | 1086 | 10485 | 26 | 5224 | 42 |
| | | | 7 | 1736 | 5680 | 944 | 2579 | 970 | 2144 | 75 | 157 | 167 | 551 | 1662 | 3077 | 67 | 6 | 165 | 1138 | 18 | 4128 | 33 |
| | m8 | 15 | 1 | 959 | 8282 | 1242 | 3336 | 5167 | 14953 | 264 | 620 | 573 | 2797 | 11258 | 8965 | 63 | 49 | 7784 | 22325 | 27 | 34615 | 77 |
| | | | 7 | 2001 | 10043 | 784 | 1959 | 7050 | 20947 | 180 | 425 | 398 | 2345 | 9777 | 7126 | 65 | 37 | 1121 | 2937 | 25 | 32270 | 72 |
| | m27 | 21 | 1 | 445 | 5485 | 1145 | 3097 | 2927 | 7256 | 257 | 586 | 569 | 1669 | 5609 | 6563 | 65 | 22 | 2164 | 12513 | 28 | 15615 | 69 |
| | | | 7 | 909 | 6651 | 708 | 1823 | 3794 | 9502 | 171 | 407 | 378 | 1227 | 4136 | 4517 | 67 | 13 | 377 | 1337 | 25 | 14422 | 64 |
| Rn | m6 | 15 | 1 | 635 | 2806 | 709 | 2546 | 1049 | 2523 | 69 | 148 | 147 | 442 | 1361 | 3040 | 65 | 10 | 617 | 6851 | 18 | 4649 | 53 |
| | | | 7 | 914 | 3370 | 379 | 1291 | 1255 | 3033 | 60 | 128 | 143 | 283 | 882 | 1640 | 66 | 5 | 93 | 509 | 17 | 4136 | 47 |

(NoG: probe sets which do not target any known gene, SS,SM,MS,MM,CX and HX : probe set classes as defined in text, M: multiple probe sets, Sp: species, Chip: our chip model name (see Table 1), NetNb: number of networks, PNb: probe number limit, P: probe set, B: Bicluster, G: gene, Occ: mean density of point in biclusters (%), %T: percentage of probe sets).

Two probe sets assigned to the same gene may have variable values for positive and negative correlation in networks constructed from different set of biological conditions.This is also true for the similarity of their neighbourhood, measured by the p-value of occurrence, under hypergeometric distribution hypothesis, of an overlap of size equal to or higher than N, given the sizes N1 and N2 of each probe set neighbourhood. We reasoned that *bona fide* similar paired probe sets should be positively correlated in all networks. A first indication of this is provided by the distribution of the frequency of paired probe sets which are positively correlated in a given number of networks. As shown in Figure 2-A, the most frequent value corresponds to the paired probe sets that are correlated in all 15 networks. A second indication is that distributions indexed on each possible number of positive networks, ranging from 1 to 14, form an ordered group of more or less regularly spaced curves, which is distinctly separate from the curve corresponding to the probe sets that are correlated in all 15 networks (arrowed curves in Figure 2-B,-C and -D). We therefore elected this special distribution to calculate the 5th percentile of positive correlation ($corr^{5th}$) and the 95th percentile of negative correlation ($anti^{95th}$) and p-value of neighbourhood overlap ($overlap^{95th}$) the limits of which were used to devise a test that classes a probe set pair j as similar in a network i if and only if:

$$_i corr^j \geq corr^{5th} \text{ and } _i anti^j \leq anti^{95th} \text{ and } _i overlap^j \leq overlap^{95th} \quad .$$

We found that these test limits were specific for each chip model, but were largely independent of the number of networks considered, which is a supplementary proof of the general validity of our method (cf. Table 3 where stability of test limits is observed in a range of 11 to 63 networks for chip m27, and in a range of 18 to 36 for chip m3). Similarly, changing biological conditions used to construct the different networks has no effect on these values, as shown with the m27 values which are derived from four different series of 21 networks (Table 3).

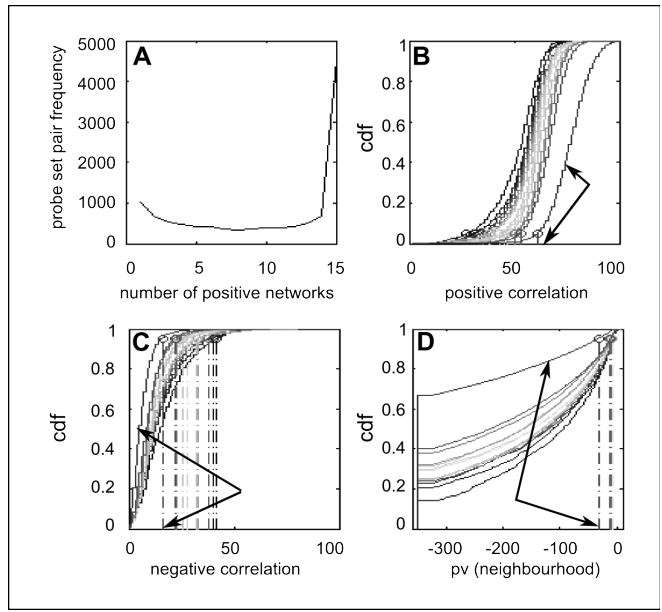

**Fig. 2. Distribution of probe set pair properties (m27-1p). A** – frequency of pairs of probe sets, belonging to class SS, which are positively correlated in one to fifteen networks. **B** – distributions of positive correlation between paired probe sets. **C** – distributions of negative correlation between paired probe sets. **D** – distributions of the neighbourhood overlap p-values of paired probe sets. In B, C and D, there are fifteen curves, one for each subset of paired probe set that are positively correlated in a given number of networks ranging from 1 to 15. The curves corresponding to the paired probe sets positively correlated in all the network and their 5$^{th}$ (B) or 95$^{th}$ (C,D) percentiles (that are used as limits to test if paired probe sets must be considered as similar in a given network) are indicated by arrowed lines.

**Table 3. Limits used to test whether paired probe sets are similar.**

| Species | Chips | NetNb | ProbeNb | CORR | ANTI | PV |
|---|---|---|---|---|---|---|
| Human | m2 | 21 | 1 | 54 | 16 | -16 |
| | | 21 | 7 | 54 | 16 | -16 |
| | m3 | 18 | 1 | 59 | 8 | -10 |
| | | 18 | 1 | 60 | 7 | -9 |
| | | 36 | 1 | 61 | 7 | -11 |
| | | 36 | 7 | 61 | 7 | -11 |
| Mouse | m5 | 35 | 1 | 62 | 8 | -6 |
| | | 35 | 7 | 62 | 8 | -6 |
| | m8 | 15 | 1 | 61 | 16 | -32 |
| | | 15 | 7 | 61 | 16 | -33 |
| | m27 | 11 | 1 | 61 | 12 | -9 |
| | | 21 | 1 | 62 | 12 | -10 |
| | | 21 | 1 | 62 | 12 | -10 |
| | | 21 | 1 | 62 | 12 | -10 |
| | | 21 | 1 | 62 | 12 | -11 |
| | | 32 | 1 | 63 | 11 | -11 |
| | | 42 | 1 | 64 | 11 | -11 |
| | | 63 | 1 | 65 | 11 | -12 |
| | | 21 | 7 | 62 | 12 | -10 |
| Rat | m6 | 15 | 1 | 59 | 8 | -17 |
| | | 15 | 7 | 59 | 8 | -17 |

Columns indicate respectively the species, the chip name, the number of networks used to calculate the limits, the minimal number of probes that a probe set must have in a gene to be considered as targeting this gene, the positive correlation limit, the negative correlation limit, and the neighbourhood overlap limit (logarithm of p-value). m27 limits for ProbeNb=1 and NetNb=21 were calculated on networks constructed on different combinations of biological conditions.

Paired probe sets which are similar in one network are not necessarily similar in another one, given the different biological conditions used to construct each network. We therefore considered that the number of networks positive for a given pair of probe sets represented the reproducibility of its similarity. We proceeded at predetermined percentage of positive networks: 1% (in fact 1 network, since less than 100 networks were used to date), 25%, 50% 75% and 100% and found, as shown in figure 3, that the frequency of similarity is an inverse linear function of the reproducibility. Contrarily to what we observed for test limits, the frequency of similarity was slightly dependent on the number of networks considered: as expected the higher the number of networks used, the higher the probability of finding a single network where a given pair of probe sets is either similar or dissimilar, which moves the frequency from 1% and 100% upwards and downwards, respectively, as observed in Figure 3-B. Reproducibility is therefore not an absolute measure, and always refers to the number of networks used to calculate it, especially for high reproducibility values.

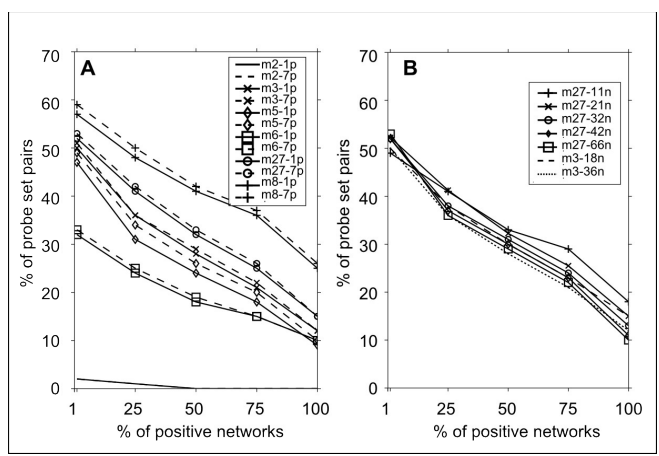

**Fig. 3. Percentage of similar paired probe sets for each level of reproducibility.** A – Effect of shifting from one to seven the minimum number of probes that a probe set must have in a gene to be considered targeting this gene. Continuous and interrupted lines indicate a minimum number of probes of one and seven, respectively. B – Effect of the number of networks used on the frequency of similar paired probe sets at different levels of reproducibility.

As we wanted to group, in each bicluster, all probe sets that target the same transcript(s) -and not simply consider each probe set pair independently-, we developed the following method. First we searched all probe set triangles built up with three pairs of similar probe sets (if there exist similar pairs (A,B), (BC) and (CD) then we generate a triangle (ABC)), and in a second step we aggregated all triangles which had a common edge. By doing so a pair of dissimilar probe sets was sometimes added (for example if triangles (ABC) and (ABD) were merged to form probe set group (ABCD), and if neither triangle (ACD) nor triangle (BCD) exists, we *de facto* introduce probe set pair CD which is not similar). We considered as acceptable to make such an exception since distribution of these added pairs shows that most of them are similar in a high number of networks (see statistics on bad links in figure 43 of supplement 1). As there are 22690 probe sets in common between m8 and m27, we were able to test the reproducibility of group assignment across different chip models, and found that only 2% of groups defined in m27 were not found in m8.

### 3.3 Comparison with Elbez results

In order to compare our results with those already published, we constructed two sets of paired probe sets that could be assimilated to 'good' and 'bad' categories defined by Elbez. (Elbez et al., 2006). More precisely, we considered as good all paired probe sets that are similar at the reproducibility level of 100%, and as bad all paired probe sets that are not similar at the reproducibility level of 1%. Elbez defined a third category grouping non informative pairs (NI in Table 3) that is pairs with at least one probe set which was considered as present or had a fold change (defined as the ratio of its signal to the mean signal calculated on the whole dataset) greater than 2 in less than 10% of the experimental points in all datasets. In our approach we did not detect such non informative probe sets, and we regrouped in an 'intermediate' category (Inter in Table 4) in which all paired probe sets that are similar up to at a maximal reproducibility level of 75%. Table 4 shows how our approach allows to qualify the results based on a binary classification by introducing the notion of reproducibility: only 22% of the paired probe sets defined as good in Elbez classification should be always considered as similar while 50% of them effectively target the same transcript(s) but only in specific circumstances. Similarly, most of the bad pairs (77%) defined in Elbez classification were classified identically by us, but 22% of them had probe sets that were similar only in some circumstances. Finally, our approach, by considering far more biological samples, allowed us to determine the nature of pairs considered as non informative by Elbez: 67% of these pairs are bad and 31% are intermediate.

**Table 4. Comparison between probe set pair classifications based either on Pearson's correlation coefficient or on PSAWN method (m3-1p).**

|  |  |  | Elbez (117222 pairs) | | |
|---|---|---|---|---|---|
|  |  |  | Good | Bad | NI |
|  |  | 10565 in common | 46 | 12 | 42 |
| PSAWN (12385 pairs) | Good | 11 | 91 \| 22 | 1 \| 1 | 8 \| 2 |
|  | Bad | 50 | 25 \| 28 | 19 \| 77 | 56 \| 67 |
|  | Inter | 39 | 60 \| 50 | 7 \| 22 | 33 \| 31 |

The third column and third line refer to the percentage of the 10565 common probe set pairs belonging to the categories defined respectively in our study and in Elbez study. Figures placed before (respectively after) the vertical lines sum up to 100 horizontally (respectively vertically), and indicate how probe sets belonging to a particular category in one study are distributed among the three categories of the other study.

### 3.4 Rank difference distributions

The method we have developed delivers an information which is probabilistic by nature (and we are aware that the probability that a given pair of probe sets is similar in e.g. 50% of networks does not mean that this pair is similar in 50% of biological conditions). However, there should exist a link between the reproducibility level of the similarity we have defined for a pair of probe sets – quantified by the percentage of positive networks – and its propensity to be similar in a given biological condition.

Indeed, if we use rank difference between two paired probe sets as an indication of their similarity in a given biological condition, we observe that the corresponding distribution curves are ordered according to the level of reproducibility, as shown in figure 4. This observation paves the way for developing new methods to assess the similarity of paired probe sets in a given experimental point, a question for which no answer exists to date.

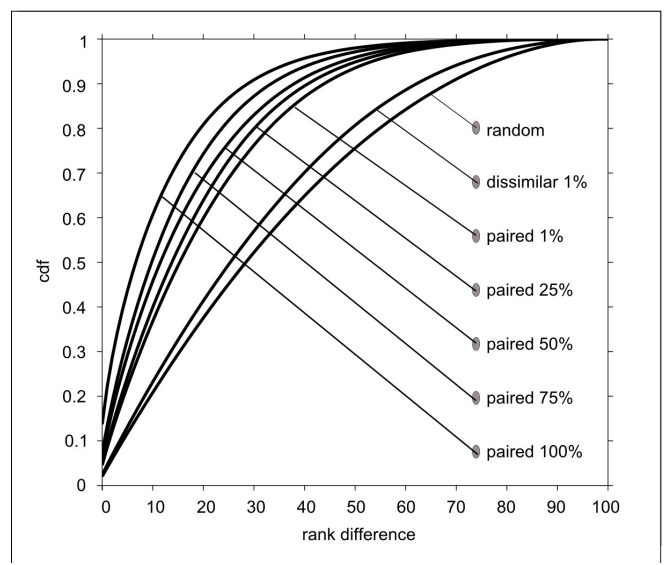

**Fig. 4. Distribution of paired probe set rank difference (m27-1p).** Paired probe sets are partitioned according to the level of reproducibility of their similarity (e.g. paired probe sets marked 'paired 25%' are similar in exactly 25% of networks). Probe set pairs which are either sampled randomly or dissimilar in at least 1% of networks are marked 'dissimilar1%' and 'random', respectively.

## 3.5 Conclusion

Our work has led to the development of a complete set of tools in Python and Matlab which allows for thorough analysis of probe set characteristics within transcriptional networks. The software we developed delivers a full textual description of each probe set (which genes, exons and transcripts are targeted by a given number of probes) and synthetic tables indicating the association between probe sets (which probe set targets the same group of transcripts with a given frequency) as well as other information on the number and properties of secondary targets (figures 12,13,14,15,54,55 and 56 of Supplement 1). These processed data, available for the most frequently referenced 3' IVT Affymetrix chips, can be used to filter out microarray results on certain properties (e.g. keeping only probe sets that target exactly one gene, with at least 7 probes, and merging signals of probe sets that target the same groups of transcripts in at least 50% of networks), which would facilitate, for example, Gene Set Enrichment Analysis (GSEA). These data will also be essential to answer related questions: which probe set could be considered as targeting the same group of transcripts in a particular microarray result? What is the correspondence between probe sets in different chip models of the same species and even across related species ?

## ACKNOWLEDGEMENTS

We thanks Azzeddine Ghers and Nicolas Rouquette for their participation to preliminary works concerning this project. We aknowledge M.C. Morris for proofreading the manuscript, and Hendrik Luuk for his comments.

*Funding*: Computing needed by this project has been realized using HPC resources from GENCI-CINES (http://www.cines.fr/) [grant number c201007503].